\documentclass[twocolumn,showpacs]{revtex4}
\usepackage{graphicx}
\usepackage{bm} 
\usepackage{amssymb} 
\usepackage{amsmath} 

\begin{document}

\title{Spin-polarized tunneling current through a thin film of a topological insulator in a parallel magnetic field}

\author{Sergey S. Pershoguba and Victor M. Yakovenko}

\affiliation{Condensed Matter Theory Center and Center for Nanophysics and Advanced Materials, Department of Physics, University of Maryland, College Park, MD 20742-4111, USA.}

\date{September 3, 2012}

\begin{abstract}
  We calculate the tunneling conductance between the surface states on the opposite sides of an ultra-thin film of a topological insulator in a parallel magnetic field. The parallel magnetic field produces a relative shift of the in-plane momenta of the two surfaces states. An overlap between the shifted Fermi circles and spinor wave functions result in unusual dependence of the tunneling conductance on the magnetic field. Because spin orientation of the electronic surface states in topological insulators is locked to momentum, spin polarization of the tunneling current can be controlled by the magnetic field.
\end{abstract}

\pacs{03.65.Vf, 73.50.Jt}

\maketitle
\section{Introduction.} \label{sec:Intro}

In the past few years, there has been a rapid progress in the field of topological insulators (TIs) \cite{TI-Review}. The three-dimensional (3D) topological insulators have the helically-spin-polarized surface states with the two-dimensional (2D) Dirac dispersion, which are observed experimentally \cite{TI-Experiments}. TIs not only offer an exciting playground for fundamental physics, but also have a variety of potential applications ranging from spintronics \cite{Garate-2010,Yokoyama-2010,Yokoyama-2010a,Burkov-2010,Mondal-2010,Zhai-2011,Wu-2011,Appelbaum-2011,Suwanvarangkoon-2011,JZhang-2012} to quantum computing \cite{QuantumComputing}. However, the experiments show that the Fermi level is often lifted to the conduction band, thus making the topological surface states less relevant for the properties of real materials \cite{BadInsulators}. In part, this motivated the study of the thin films of TIs, where the Fermi level, bulk gap and the hybridization between the opposite surface states can be tuned \cite{ThinFilmsTuning,FieldEffectTransistor,Molenkamp}.  A number of intriguing effects for the TI films were predicted in the case where the top and bottom surfaces interact \cite{TI-effects}. In the ultra-thin limit, a gap opens in the surface states electronic spectrum due to hybridization between the opposite surfaces \cite{FieldEffectTransistor,ThinFilmModels,YZhang-2010,Taskin-2012}. 

While most papers study the in-plane transport properties of the surface states, here we study the tunneling conductance $G$ per unit area across the ultra-thin TI film in the presence of a parallel magnetic field $B_y$ (see Fig.~\ref{fig:Film}). The in-plane magnetic field results in a relative shift of the Dirac cones of the surface states in momentum space, as discussed in related works on graphene bilayer \cite{Parallel1,Parallel2}. However, unlike the spectrum of graphene, the spectrum of the TI thin film remains gapped until a critical value of the magnetic field is achieved \cite{Zyuzin-2011}. Motivated by the original measurements of $G$ vs $B_y$  for a GaAs bilayer by Eisenstein et al. \cite{Eisenstein-1991}, here we calculate the tunneling conductance for a TI film as a function of the parallel magnetic field. In contrast to other systems, the spin structure of the electronic spectrum for the surface states of TI results in unusual dependence of the tunneling conductance on $B_y$. In a recent experiment \cite{Parallel-experiment}, the authors observed a strong $B_y$ dependence of the out-of-plane current in a film of Bi$_2$Te$_3$.  The predicted theoretical dependence $G$ vs $B_y$ is in qualitative agreement with the experimental curve. In addition, we predict that the spin polarization of the tunneling current is proportional to the magnetic field. The $100 \%$ spin polarization can be achieved for an experimentally accessible strength of the magnetic field.

\section{The Model.}
\begin{figure}
\includegraphics[width=0.6\linewidth]{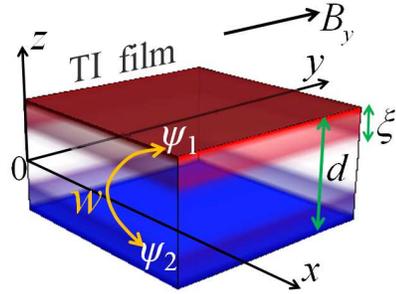}
\caption{(Color online) Thin film of a topological insulator of thickness $d$ in a parallel magnetic field $B_y$. The surface states $\psi_1$ and $\psi_2$ shown in red and blue overlap and couple when the thickness $d$ is comparable with the decay length $\xi$ of the surface state.} \label{fig:Film}
\end{figure}
Let us consider a film of TI, which has two states $\psi_1$ and $\psi_2$ localized at the opposite surfaces of the film, as shown in red and blue in Fig.~\ref{fig:Film}. While the wave functions are localized in the $z$ direction, electrons are free to move parallel to the surface, so the in-plane momentum $\bm p = (p_x,p_y)$ is a good quantum number. For clarity of consideration, we model the surface states by the simple Rashba Hamiltonians 
\begin{eqnarray}
  && h_{\psi_1,\psi_2} = \pm h(\bm p),\label{h12} \\
  && h(\bm p) = v\,\hat{\bm z}(\hat{\bm \sigma}\times\bm p)= v(\hat \sigma_xp_y-\hat \sigma_yp_x).\label{h}
\end{eqnarray}
Different signs $\pm$ correspond to the surface states $\psi_1$ and $\psi_2$ and describe the unit Rashba vectors $\pm \hat {\bm z}$, which are collinear with the normals to the corresponding surfaces. The variable $v$ has dimensions of velocity, and $\hat \sigma_{\alpha}=(\hat 1,\hat{\bm \sigma})$ denotes a full set of the operators acting in the spin space, where $\hat 1$ is a $2\times2$ unit matrix and $\hat{\bm \sigma}=(\hat \sigma_x,\hat \sigma_y,\hat \sigma_z)$ are the Pauli matrices. Hamiltonian~(\ref{h}) has the following eigenstates and eigenenergies
\begin{equation}
  \mid\bm p,\pm\rangle = \frac{1}{\sqrt 2}
  \left(
  \begin{array}{c}
    1 \\
    \pm e^{-i\arctan(p_x/p_y)}
  \end{array}
  \right),\,\,
  E_0(\bm p) = \pm v|\bm p|,    \label{heigenState}
\end{equation}
where the spectrum has a well-known form of the 2D Dirac cone. The spinors in Eq.~(\ref{heigenState}) describe the helical spin-momentum locking with the spin polarization perpendicular to the momentum
\begin{equation}
	\bm S(\bm p) = \langle \bm p,\pm\mid \hat{\bm \sigma} \mid \bm p,\pm \rangle =\pm\frac{(\bm p\times \hat{\bm z})}{|\bm p|}, \label{helical}
\end{equation}
so the helicity~(\ref{helical}) is opposite for the $\mid \bm p,+ \rangle$ and $\mid \bm p,- \rangle$ bands. Because of the minus sign for $h_{\psi_2}$ in Eq.~(\ref{h12}), the bands $\mid \bm p,\pm \rangle$ are energetically-inverted, and, so, the helicities for the opposite surface states $\psi_1$ and $\psi_2$ are opposite for a fixed energy. The wave functions of the surface states $\psi_1$ and $\psi_2$ decay into the bulk and have a finite decay length $\xi$, as illustrated in Fig.~\ref{fig:Film}.  So, when the thickness of the film $d$ becomes comparable with the decay length $d\sim\xi$, there is a finite coupling between the surface states $w $, which we assume to be proportional to the unit matrix in the spin space. So, the TI film can be modeled by a $4\times4$ Hamiltonian $H(\bm p)$, which acts on the 4-component wave function $\Psi$
\begin{eqnarray}
 && H(\bm p) = \left(
           \begin{array}{cc}
              h(\bm p) & w \\
              w & -h(\bm p) \\
	   \end{array} \right),\quad
    \Psi = \left(
           \begin{array}{c}
             \psi_1\\
	     \psi_2
	   \end{array} \right). \label{Ham}
\end{eqnarray}
This Hamiltonian was also used in Ref.~\cite{Zyuzin-2011} for description of a thin film. Even though the bulk bands usually contribute to the electronic properties of the real materials \cite{BadInsulators}, we ignore a possible bulk effect in order to highlight the contribution of the surface states. Hamiltonian~(\ref{Ham}) has the spectrum $E(\bm p) = \pm \sqrt{v^2p^2+w^2}$, where the energy gap is determined by the tunneling element $w$. Experimentally, gap varies from $0.25$~eV for the ultra-thin $2$~nm film to $0.05$~eV for the $5$~nm film of Bi$_2$Se$_3$~\cite{YZhang-2010}.

\section{Spectrum in a parallel field.}
\begin{figure}
  \includegraphics[height=0.6\linewidth]{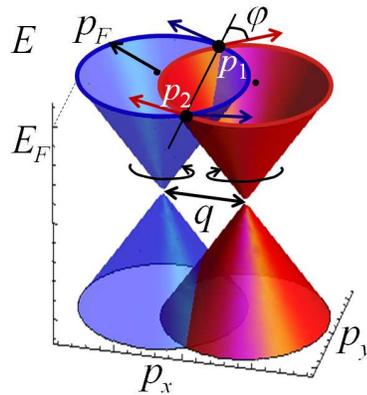}
  \caption{(Color online) Electronic spectrum~(\ref{specParallel}) of the thin film consists of the two Dirac cones spin-polarized in the opposite directions as shown with circulating arrows. The Dirac cones are shifted by $q_x$ due to the parallel magnetic field $B_y$. Tunneling between the Dirac cones is dominated by the electrons with momenta $\bm p_1$ and $\bm p_2$ where the shifted Fermi circles $|\bm p\pm\bm q/2|=p_F$ intersect.} \label{fig:shiftedCones}
\end{figure}

Let us now discuss the spectrum of the TI thin film in a parallel magnetic field $\bm B = B_y\hat{\bm y}$. In the Landau gauge $\bm A = zB_y\,\hat{\bm x}$, the Peierls substitution $\bm p\rightarrow\bm p-e\bm A$ transforms the Hamiltonian~(\ref{Ham}) into
\begin{equation}
 H(\bm p) = \left(
           \begin{array}{cc}
	     h(\bm p-\bm q/2) & w \\
	     w & -h(\bm p+\bm q/2) \\
	   \end{array} \right),\,\, \bm q =  eB_yd\,\hat{\bm x}. \label{Hamq}
\end{equation}
The Zeeman term can be included in the definition of $q$ \cite{Zyuzin-2011,Zeeman}. Notice that the parallel magnetic field leads to the opposite momentum shifts $\Delta p_x=\pm q/2$ on the opposite surfaces located at $z=\pm d/2$. Indeed, consider an electron traveling between the surfaces with the classical velocity $v_z$ in the $z$ direction. Then, the Lorentz force $\bm F_L=ev_zB_y\hat{\bm x}$ results in the in-plane momentum change $\bm q = \int dt\bm F_L =  eB_yd\,\hat{\bm x}$ between the opposite surfaces. So, in the limit $w\rightarrow 0$, the spectrum of the Hamiltonian~(\ref{Hamq}) is given by the two shifted Dirac cones
\begin{equation}
     E(\bm p) = \pm v\left|\bm p \pm\frac{\bm q}{2}\right|, \label{specParallel}
\end{equation}
 as shown in Fig.~\ref{fig:shiftedCones}. A similar shift of the Dirac cones was discussed for graphene multilayer due to a parallel magnetic field \cite{Parallel1,Parallel2} and for a twisted graphene bilayer \cite{Andrei-2010}. In general, when $w$ is not small, the spectrum of Hamiltonian~(\ref{Hamq}) was calculated in Ref.~\cite{Zyuzin-2011}. However, we focus on the former case when $w$ is small and include $w$ as a perturbation.

\section{Tunneling current in a parallel field.}
\begin{figure}
\begin{tabular}{c}
    \includegraphics[width=0.6\linewidth]{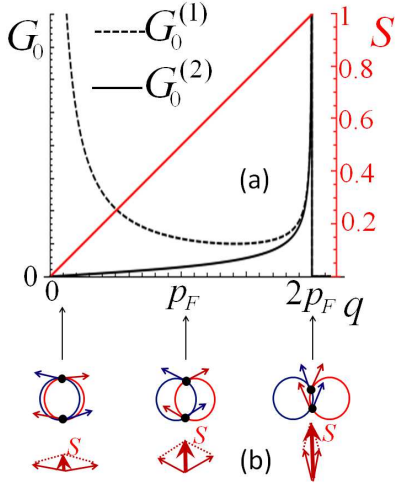}
\end{tabular}
\caption{(Color online) (a) The tunneling conductances for spin-unpolarized~$G_0^{(1)}$, Eq.~(\ref{sigmaParallel0}), and spin-polarized~$G_0^{(2)}$, Eq.~(\ref{sigmaParallel1}), Fermi circles are shown by the dashed and solid black lines, respectively. The latter curve $G_0^{(2)}$ corresponds to the negative magnetoresistance. The red curve is the spin polarization~(\ref{spinPolarization}) of the tunneling current, and the right vertical axis shows its value, where $1$ corresponds to $100\%$ spin polarization. (b) Schematic views of the shifted Fermi circles $|\bm p\pm\bm q/2|=p_F$ for the corresponding values of $q$. The tunneling is allowed for the momenta $\bm p_1$ and $\bm p_2$ where the Fermi circles intersect, also shown in Fig.~\ref{fig:shiftedCones}. The spin polarizations for these momenta are shown by the blue and red arrows corresponding to the $\psi_1$ and $\psi_2$ surfaces. The vector sum of the arrows of the same color defines the net spin polarization of the tunneling current, which grows with the increase of $q$ as illustrated in the bottom of panel (b).}  \label{fig:transport}
\end{figure}

When a small potential difference $V$ is applied between the opposite surfaces of the film, the out-of-plane tunneling current per unit area
\begin{equation}
  j_\alpha =(j_0,\bm j)=  iw\left(e^{ieVt}\psi_1^\dag \hat\sigma_\alpha\psi_2-e^{-ieVt}\psi_2^\dag \hat\sigma_\alpha\psi_1\right)
  \label{tunnelcur}
\end{equation}
flows between the opposite surfaces, where $\psi_1$ and $\psi_2$ are the annihilation operators on the corresponding surface states. Here, we put $\hbar=1$ for simplicity and reestablish the correct units in the end of calculation.  So, the tunneling current is written in a convenient four-component form, such that $e j_0$ and $\frac{\hbar}{2}\bm j$ are the charge and spin currents. Using the linear-response theory for $w$ as a perturbation \cite{TunnelingConductance}, the tunneling conductances per unit area $G_\alpha=\lim_{V\rightarrow 0}{\langle j_\alpha\rangle}/{V}$ can be expressed through the spectral functions $A_{1,2}(E,\bm p)$ evaluated at the Fermi energy $E_F$ as
\begin{equation}
	G_\alpha = \frac{w^2e}{2\pi}\int d^2 p\, {\rm Tr}\left[\hat \sigma_\alpha A_1\left(E_F,\bm p+ \frac{\bm q}{2}\right) A_2\left(E_F,\bm p-\frac{\bm q}{2}\right) \right]. \label{sigmaGeneral}
\end{equation}
Here, $ G_\alpha=( G_0,\bm G)$ has four components, where $e G_0$ and $\frac{\hbar}{2}\bm G$ correspond to the tunneling conductance for charge and spin currents. The spectral functions $A_{1,2}$ are given in the momentum space
\begin{eqnarray}
	A_{1,2}\left(E_F,\bm p\pm\frac{\bm q}{2}\right) = && \delta(v|\bm p\mp\frac{\bm q}{2}|-E_F)  \label{momentumSpace}\\ 
      && \times\mid\bm p\mp\frac{\bm q}{2},\pm\rangle \langle \bm p \mp\frac{\bm q}{2},\pm\mid, \nonumber  
\end{eqnarray}
where the  opposite signs correspond to the opposite surfaces $\psi_1$ and $\psi_2$, and the spinors $\mid \bm p,\pm\rangle$ are defined in Eq.~(\ref{heigenState}). Using Eqs.~(\ref{sigmaGeneral})~and~(\ref{momentumSpace}), let us first calculate the tunneling conductance $G_0$ for the charge current
\begin{eqnarray}
 &  G_{0} = \frac{w^2e}{2\pi v^2} \int d^2p\, \delta\left(\left|\bm p - \frac{\bm q}{2} \right|-p_F\right) \delta\left(\left|\bm p + \frac{\bm q}{2} \right|-p_F\right) f(\bm p),\nonumber\\
 &  f(\bm p)=\left|\langle\bm p-\frac{\bm q}{2},+ \mid\bm p+\frac{\bm q}{2},-\rangle\right|^2,\quad \label{sigmaParallel}
\end{eqnarray}
where $p_F=E_F/v$ is the Fermi momentum. According to Eq.~(\ref{sigmaParallel}), the tunneling current is carried by the electrons that have the in-plane momenta $\bm p_1$ and $\bm p_2$ corresponding to the intersection points of the shifted Fermi circles $|\bm p\pm\bm q/2|=p_F$, as shown by the thick dots in Fig.~\ref{fig:shiftedCones}. In addition, there is a contribution $f(\bm p)$ due to the non-trivial scalar product between the spinors~(\ref{heigenState}) corresponding to the Fermi circles.

As a warm up, let us first consider the case where is no spin-momentum locking, so $f(\bm p)=1$ and Eq.~(\ref{sigmaParallel}) gives
\begin{equation}
   G_{0}^{(1)}(q)=\frac{2w^2ep_F^2}{\pi\hbar^3 v^2q\sqrt{4p_F^2-q^2}}. \label{sigmaParallel0}
\end{equation}
The tunneling conductance $ G_{0}^{(1)}$ diverges at $q=0$ and $q=2p_F$, as shown by the dashed line in Fig.~\ref{fig:transport}(a). For a small magnetic field $q\ll 2p_F$, the tunneling conductance is large because of the large overlap between the two Fermi circles shown in the left part of Fig.~\ref{fig:transport}(b). For an intermediate magnetic field $q\sim p_F$, the Fermi circles intersect only at two points $\bm p_1$ and $\bm p_2$, so the tunneling conductance decreases. For $q\sim 2p_F$, the two points $\bm p_1$ and $\bm p_2$ come together, so the Fermi circles overlap at the locally flat regions as shown in the right part of Fig.~\ref{fig:shiftedCones}(b), and the tunneling conductance is large again. Once $q>2p_F$, the two Fermi circles separate, and the tunneling conductance drops to zero. The experimental curve of the tunneling conductance for a GaAs bilayer \cite{Eisenstein-1991} shows behavior similar to Eq.~(\ref{sigmaParallel0}), but the divergences at $q=0$ and $q=2p_F$ are smeared out due to a finite scattering time $\tau$.

Now, let us consider the form-factor $f(\bm p)$ in Eq.~(\ref{sigmaParallel}) arising from the spinor eigenstates~(\ref{heigenState}) of the Rashba Hamiltonian~(\ref{h}). As discussed above, the surface states have opposite spin polarizations. In Figs.~\ref{fig:shiftedCones} and \ref{fig:transport}, the polarizations corresponding to the different Fermi circles are shown by the blue and red arrows intersecting at the angle $\phi$, the value of which follows from simple trigonometry: $\cos\phi = q/2p_F$. The form-factor $f(\bm p)$ in Eq.~(\ref{sigmaParallel}) takes into account an overlap between the spinors on the different Fermi circles and is equal to $f(\bm p_{1,2})=  \cos^2\phi$. So, the tunneling conductance~(\ref{sigmaParallel}) is multiplied by a factor of $q^2/4p_F^2$ relative to Eq.~(\ref{sigmaParallel0})
\begin{equation}
   G_{0}^{(2)}(q)=\frac{w^2eq}{2\pi\hbar^3v^2\sqrt{4p_F^2-q^2}}. \label{sigmaParallel1}
\end{equation}
The tunneling conductance given by Eq.~(\ref{sigmaParallel1}) is plotted by the solid line in Fig.~\ref{fig:transport}(a). In contrast to $G^{(1)}_{0}$, the tunneling conductance $ G^{(2)}_{0}$ is suppressed as $q\rightarrow 0$, because spin polarizations of the Fermi circles are opposite at the points $\bm p_1$ and $\bm p_2$, as shown in Fig~\ref{fig:transport}(b) on the left. This is a signature of the opposite spin helicity of the Fermi circles and thus may be used as a way to detect it. For $q$ near $2p_F$, however, the spin configurations are almost aligned, as shown in Fig.~\ref{fig:transport}(b), so, $ G_{0}^{(1)}$ and $ G_{0}^{(2)}$ become equal.

The non-trivial spin structure of the Fermi circles not only modifies the tunneling current, but also makes it spin-polarized. Let us define spin polarization as the ratio of the spin current over the charge current
\begin{equation}
  \bm S=\frac{\langle\bm j\rangle}{\langle j_0\rangle}=\frac{\bm  G}{ G_0}, \label{spinPolarization0}
\end{equation}
where the second equation follows from the definition of the conductance. Using the property that the eigenstates~(\ref{heigenState}) have the well-defined spin orientation~(\ref{helical}), we obtain from Eq.~(\ref{sigmaGeneral})
\begin{equation}
\bm S = \frac{\bm S(\bm p_1-\bm q/2)+\bm S(\bm p_2-\bm q/2)}{2}=\cos \phi\,\,\hat{\bm y}=\frac{ed}{2p_F}\,\bm B.
  \label{spinPolarization}
\end{equation}
So, the spin polarization of the tunneling current is determined by the vector sum of the spin directions at the momenta $\bm p_1$ and $\bm p_2$. Indeed, in the process of tunneling, the electrons with the in-plane momenta $\bm p_1$ and $\bm p_2$ move from the ``blue'' to ``red'' Fermi circle, carrying the total spin, which is a vector sum of the spins at the momenta $\bm p_1$ and $\bm p_2$, in agreement with Eq.~(\ref{spinPolarization}). The helical spin configuration~(\ref{helical}) is such that the net spin polarization $\bm S$ is parallel and proportional to $\bm B$, as illustrated in Fig.~\ref{fig:transport}(b). At $q=0$, the spins at the points $\bm p_1$ and $\bm p_2$ are opposite, and the spin polarization of the tunneling current vanishes. At $q=2p_F$, the spins at the points $\bm p_1$ and $\bm p_2$ are collinear and the tunneling current is fully spin-polarized. Also, notice that the electron tunneling at $\bm p_1$ changes the spin polarization from the ``blue'' to ``red'' arrow, thus creating a torque $\bm S(\bm p_1+\bm q/2)-\bm S(\bm p_1-\bm q/2)$ in $\hat {\bm x}$ direction. However, the torque is opposite at $\bm p_2$, so the net torque is zero.

\section{Experimental relevance.}

Let us estimate the critical magnetic field $B_y$ where the Fermi circles almost detach, i.e., where $q=2p_F$. We take the realistic value $v = 5\times 10^5$~m/s, assume $d=5$~nm (5 quintuple layers of Bi$_2$Se$_3$) and carrier concentration corresponding to $E_F = 30$~meV. Using Eq.~(\ref{Hamq}) for $q$, we estimate the critical value of the field as $B_y=20$~T. This value is experimentally accessible and can be further reduced by either increasing the thickness $d$ or decreasing the Fermi momentum $p_F$. 

An intriguing strong negative magnetoresistance effect was reported for the Sn-doped films of Bi$_2$Te$_3$ in Ref.~\cite{Parallel-experiment}. A weak in-plane magnetic field less than 1 T causes a large drop  of the out-of-plane resistance $R_{zz}$ as shown in Fig.~4(c) of Ref.~\cite{Parallel-experiment}. Since $R_{zz}\propto 1/G_0$, a decrease in resistance $R_{zz}$ corresponds to an increase in conductance $G_0$, which qualitatively agrees with plot of $G^{(2)}_0$ vs $q$ shown in Fig.~\ref{fig:transport}(a) of our paper. In addition, the effect of Sn doping was studied in Ref.~\cite{Parallel-experiment}. The Sn doping leads to a decrease of carrier concentration and $p_F$, which is corroborated by an increase of resistance $R_{zz}$ in the experiment. At the same time, the magnitude of negative magnetoresistance increases dramatically with doping, as shown in Fig.~4(a,b) of Ref.~\cite{Parallel-experiment}. This observation is consistent with Eq.~(\ref{sigmaParallel1}), where the conductance $G^{(2)}_0$ increases when $p_F$ decreases (for $q<2p_F$).  However, our idealized model may be not fully applicable to the experiment \cite{Parallel-experiment}, where polycrystalline films were studied. Grain boundaries and defects may host topological states \cite{Defects,Defects1}, which can  contribute considerably to the tunneling current.

Equation~(\ref{sigmaParallel1}) was obtained for an idealized situation where the temperature $T$ and the inverse scattering time $1/\tau$ are much lower than the Fermi energy $E_F$. For realistic TI materials, the scattering rate $1/\tau$ is finite due to impurities or other mechanisms.  When $1/\tau$, $T$, and fluctuations of chemical potential \cite{Paddles} become comparable with $E_F$, the predicted effects would be smeared out.  However, the majority of TIs are not strongly-correlated systems, so the effects of interactions between electrons should not alter the predicted effects considerably. 

While the isotropic Dirac cone approximation is valid for small energies in Eq.~(\ref{heigenState}), it is not exact for higher energies, where energy spectrum has hexagonal warping anisotropy, and spin polarization has an out-of-plane component. In this case, the analytic formula~(\ref{sigmaParallel1}) is not applicable, but can be easily generalized. The appropriate spectrum $E(\bm p)$ and spinors $\mid\bm p,\pm\rangle$ should be substituted into Eq.~(\ref{momentumSpace}), and the tunneling conductance obtained from Eq.~(\ref{sigmaGeneral}). In general, the tunneling conductance may depend on the in-plane orientation of the magnetic field $\bm B$ due to anisotropy of the surface state dispersion.  This effect can be utilized to obtain information about hexagonal warping experimentally.

\section{Conclusions.} 
We have theoretically studied tunneling conductance between the opposite surface states in a thin film of TI when a parallel magnetic field is applied. The helical spin polarization and the overlap between the Fermi circles result in the unusual spin polarization of the tunneling current. Our theoretical results are qualitatively consistent with the experiment \cite{Parallel-experiment}. However, further experimental verifications are needed. First, magnetoresistance in Ref.~\cite{Parallel-experiment} was measured for a relatively weak magnetic field $B<1$~T. We predict {\it non-monotonous} behavior of magnetoresistance for a large magnetic field: the conductance (resistance) could sharply decrease (increase) when the magnetic field exceed the condition $B>2p_F/ed$. Second, a measurement of the spin polarization of the tunneling current is desirable. This effect may be important for spintronic applications and may pave the way to  observation of spin-polarized currents in TIs.



\begin{thebibliography}{99}
\bibitem{Zeeman} In the presence of the Zeeman term the magnetic shift becomes $q = B_y(ed-2g\mu_B /v)$, where $\mu_B$ is the Bohr magneton, $v=  5\times10^5$~m/s, and $g \sim 1$ is the $g$-factor. The Zeeman contribution is comparable to the orbital contribution only for a small thickness $d =2g\mu_B/ev\sim 0.2$ nm. So, the Zeeman contribution can be neglected for the films used in experiments, where $d> 1$~nm.
\end{thebibliography}
\end{document}